\documentclass[12pt]{article}
\usepackage{epsfig}
\newcommand{\ket}[1]{|#1\rangle}
\newcommand{\bra}[1]{\langle #1|}

\title{On the alleged nonlocal and topological nature of the 
molecular Aharonov-Bohm effect}
\author{Erik Sj\"{o}qvist\footnote{Electronic address: 
erik.sjoqvist@kvac.uu.se} \\ 
Department of Quantum Chemistry, Uppsala University, \\  
Box 518, Se-751 20 Uppsala, Sweden} 
\begin{document}
\maketitle
\begin{abstract}
The nonlocal and topological nature of the molecular
Aharonov-Bohm (MAB) effect is examined for real electronic
Hamiltonians. A notion of preferred gauge for MAB is suggested. 
The MAB effect in the linear $+$ quadratic $E\otimes \varepsilon$
Jahn-Teller system is shown to be essentially analogues to an 
anisotropic Aharonov-Casher effect for an electrically neutral 
spin$-\frac{1}{2}$ particle encircling a certain configuration 
of lines of charge.
\end{abstract} 
\newpage 
\section{Introduction}
The molecular Aharonov-Bohm (MAB) effect, first hinted 
at by Longuet-Higgins and coworkers 
\cite{longuet58,longuet61,herzberg63,longuet75}, is one of the 
paradigmatic examples on early anticipations of Berry's discovery 
\cite{berry84} of geometric phase factors accompanying cyclic 
adiabatic changes. The importance of MAB ranges from testable 
shifts of the vibronic energy spectrum \cite{kendrick97,vonbusch98},  
effects on cross-sections in molecular reactions 
\cite{kupperman93,kendrick96,adhikari00}, and effects on reduction 
factors \cite{sjoqvist94}, to subtle symmetry 
assignments of vibronic states in Jahn-Teller systems 
\cite{ham87,ham90,rios96,moate96} and search for conical intersections 
\cite{longuet75,stone76,varandas79,xantheas90,ceotto00,johansson03}. 
However, it is probably fair to say that interest in the MAB effect 
itself arose first after it was realized \cite{mead79,mead80} its 
mathematical analogy with the standard Aharonov-Bohm (AB) effect 
\cite{aharonov59}. This mathematical analogy is the independence 
of details of the shape of the closed path in a multiply connected 
region of nuclear configuration space.

In a recent work \cite{sjoqvist02}, the present author has considered
the physical nature of this analogue in the case of the $E\otimes
\varepsilon$ Jahn-Teller system. The question addressed there was: does
MAB share all the remarkable properties of AB? The main outcome of
this analysis was that although MAB obeys the above mentioned
independence of details of the closed path in nuclear configuration
space, it does not share the remaining nonlocal and topological
properties of AB. The purpose of the present paper is to develop the
local and nontopological nature of MAB further. In particular, we wish
to extend \cite{sjoqvist02} to any molecular system with real 
electronic Hamiltonians. We also wish to propose a notion of preferred
gauge for MAB in this case, in terms of a so-called gauge invariant
reference section \cite{pati95a,pati95b} that appears naturally in the
theory of the open path Berry phase
\cite{samuel88,mukunda93}.

The starting point for the argumentation in \cite{sjoqvist02} is a 
set of criteria for a nonlocal and topological phase effect, first 
explicitly introduced by Peshkin and Lipkin \cite{peshkin95}. These 
are:  
\begin{itemize}
\item A phase effect is nonlocal if:
\begin{itemize}
\item[(N1)] the system experiences no physical field,   
\item[(N2)] no exchange of physical quantity takes place along the 
system's path. 
\end{itemize}
\item A phase effect is topological if: 
\begin{itemize}
\item[(T1)] it requires the system to be confined to a
multiply connected region, 
\item[(T2)] any assignment of phase shift along the system's path 
is necessarily gauge dependent and thus neither objective nor 
experimentally testable.
\end{itemize}  
\end{itemize}  
One can argue that AB fulfills all these criteria for a nonlocal and
topological phase effect. On the other hand, in the case of MAB, where
the physical system consists of the nuclear configuration and a set
electronic variables, the situation is very different. First, MAB is
local in that it obeys neither (N1) nor (N2): the nuclei experiences
the Coulomb field from the electrons, and there must be a local
exchange of force to create the change in the electronic state
necessary for the appearance of points in nuclear configuration space,
across which the electronic open path Berry phase factor changes
sign discontinuously \cite{sjoqvist97,garcia98,englman99,englman00}.
Secondly, MAB is nontopological in that (T2) fails since there is an
objective and experimentally testable open path electronic Berry phase
that causes the effect on the nuclear motion. Only (T1) is fulfilled
for MAB: the nuclei must be confined to a multiply connected region
for the effect to occur. We wish to discuss the above criteria in
relation to MAB systems of general kind.

In the next section, we extend the argumentation of \cite{sjoqvist02}
to the case of arbitrary molecular systems with real electronic
Hamiltonians. A notion of preferred gauge for MAB in this case is 
suggested in section III and applied in detail to the $E\otimes
\varepsilon$ Jahn-Teller system. The analogue between the MAB effect in
the $E\otimes \varepsilon$ Jahn-Teller system and the Aharonov-Casher
effect \cite{aharonov84}, as pointed out in \cite{sjoqvist02}, is 
further developed in section IV, so as to treat linear and quadratic 
coupling simultaneously. The paper ends with the conclusions.

\section{Locality and topology}
The argumentation in \cite{sjoqvist02} that MAB is essentially a local
and nontopological effect was put forward in the special case
of the $E\otimes \varepsilon$ Jahn-Teller system. This raises the
question whether the main conclusions arrived at in
\cite{sjoqvist02} also apply to other molecular systems that may be 
described accurately by real electronic Hamiltonians. In this section
we address this issue and argue that MAB also in the general case
is local and nontopological.

For sake of clarity, we focus on the motion in some pseudorotational 
(internal) nuclear coordinate $\theta$, as described by the vibronic 
Hamiltonian 
\begin{eqnarray}
H = \frac{1}{2} p_{\theta}^{2} + H_e (\theta) , 
\label{eq:vibronicham}
\end{eqnarray}
where $p_{\theta}$ is the canonical momentum corresponding to $\theta$
and $H_e (\theta)$ is the electronic Hamiltonian assumed to be real,
traceless, and fulfilling $H_e (\theta+2\pi)=H_e (\theta)$.  Let $\{
\ket{n} \}_{n=1}^{N}$ be a fixed orthonormal basis of the electronic
Hilbert space of finite dimension $N$. Furthermore, let $\{ R(\theta)|
\theta \in [0,2\pi)$ be a one-parameter set of members of the rotation
group SO(N) in ${\bf R}^N$, in terms of which the orthonormal
instantaneous eigenvectors of $H_e (\theta)$ read $\{ \ket{n(\theta)} = 
R(\theta) \ket{n} \}_{n=1}^{N}$. Then, we may write
$H_e (\theta) = R(\theta) E(\theta) R^{\textrm{T}} (\theta)$ with
${\textrm{T}}$ being transpose and $E(\theta) = {\textrm{diag}} \big[
E_1(\theta),\ldots, E_N (\theta) \big]$, where $E_1
(\theta),\ldots,E_N(\theta)$ are the electronic eigenenergies
corresponding to $\ket{1(\theta)}, \ldots ,
\ket{N(\theta)}$.  

The Born-Oppenheimer regime is attained when the electronic
eigenenergies $\{ E_{n}(\theta) \}_{n=1}^N$ are well separated so that
the nuclear motion takes place on a single electronic potential energy 
surface, let us say $E_n (\theta)$. This may be described by the 
effective nuclear Hamiltonian
\begin{eqnarray} 
H_{n} = \langle n(\theta) |H| n(\theta) \rangle = 
\frac{1}{2} p_{\theta}^{2} + E_{n} (\theta) , 
\end{eqnarray}
where we have used that $\bra{n} R^{\textrm{T}} (\theta) p_{\theta}
R(\theta) \ket{n} = 0$. The issue of MAB arises when considering the
single-valuedness of the molecular state vector $\ket{\Psi (\theta)}$,
which in the Born-Oppenheimer regime is a product of the instantaneous
electronic energy eigenvector $\ket{n(\theta)}$ and a nuclear factor
$\chi(\theta)$, i.e., $\ket{\Psi (\theta)} = \chi (\theta)
\ket{n(\theta)}$. Here, we use the position representation of the
nuclear motion in the pseudorotational angle $\theta$. The
single-valuedness of $\ket{\Psi(\theta)}$ requires that any
multi-valuedness of the electronic part must be compensated for by the
nuclear part.  Thus, if $\ket{n(\theta + 2\pi)} = -\ket{n(\theta)}$, 
$\chi (\theta)$ must fulfill the boundary condition 
$\chi (\theta + 2\pi) = -\chi (\theta)$.
A nontrivial MAB effect arises if and only if there is no 
phase transformation of the form $\ket{n(\theta)} \rightarrow 
\ket{\widetilde{n}(\theta)} = e^{i\xi (\theta)} \ket{n(\theta)}$ 
so that the effective vector potential 
\begin{eqnarray}
A_n (\theta) \equiv 
i \bra{\widetilde{n}(\theta)} \partial_{\theta} 
\ket{\widetilde{n}(\theta)} = - \partial_{\theta} \xi (\theta) 
\label{eq:mabpotential}
\end{eqnarray}
vanishes everywhere and at the same time retaining single-valuedness 
for $\chi (\theta)$ under the requirement that $\ket{\Psi(\theta)}$ 
is single-valued.

Under the condition that there is a nontrivial MAB effect, let us 
ask: what is its physical nature in relation to the standard AB 
effect? Let us first address the issue of locality. As already 
indicated, the standard AB effect, which may occur when a charged 
particle encircles a line of magnetic flux, is nonlocal as it fulfills 
the criteria (N1) and (N2): the effect arises although the particle 
experiences no physical field and no exchange of physical quantity 
takes place along the particle's path. Could the same be said 
about the MAB effect? One way to address this question is to 
note that since the electronic Born-Oppenheimer states are eigenstates 
of $H_e (\theta)$, it may be tempting to replace the electronic 
motion by the appropriate eigenvalue of $H_e (\theta )$ in 
the Born-Oppenheimer regime so that the electronic variables can 
be ignored, creating an illusion that the nontrivial effect of the 
MAB vector potential on the nuclear motion is nonlocal and topological 
in the sense of the standard AB effect. However, this argument  
fails essentially because the electronic variables are 
dynamical and do not commute among themselves. Thus, there always 
exist a subset of these variables, namely those that are off-diagonal 
in the instantaneous electronic eigenbasis $\{ \ket{n(\theta)} 
\}_{n=1}^{N}$, whose expectation values vanish in the Born-Oppenheimer 
limit, but whose fluctuations do not. These fluctuations are due to 
the local interaction between the nuclear variables and the electronic 
degrees of freedom. Thus, both (N1) and (N2) fail for MAB.

Next, we address the issue of topology. The standard AB effect is
topological in that (T1) it may occur although the region of magnetic 
field is inaccessible to the encircling charged particle and in that
(T2) the charged particle only feels the gauge dependent vector 
potential along the path. In the case of MAB, recall that two types 
of degrees of freedom are involved: those of the electrons and
those associated with the nuclear configuration. Now, any assignment
of nuclear phase for open paths in nuclear configuration space is
necessarily gauge dependent and therefore unphysical. On the other
hand, there is an objective way to relate the origin of the MAB phase 
effect locally in nuclear configuration space using the open path Berry 
phase $\gamma_{n}$ for the corresponding electronic Born-Oppenheimer 
state vector $\ket{\widetilde{n} (\theta)} = e^{i\xi (\theta)} 
\ket{n(\theta)}$. Here, we assume $\xi (\theta)$ to be differentiable 
along the path but otherwise arbitrary. The noncyclic Berry phase is
defined by removing the accumulation of local phase changes from the
total phase and is testable in polarimetry \cite{garcia98,larsson03}
or in interferometry \cite{wagh98,sjoqvist01}. We obtain for
$\ket{\widetilde{n} (\theta)}$
\begin{eqnarray} 
\gamma_{n} & = & 
\arg \langle \widetilde{n}(\theta_0)\ket{\widetilde{n}(\theta)} + 
i \int_{\theta_{0}}^{\theta} \langle \widetilde{n}(\theta') | 
\frac{\partial}{\partial \theta'} |\widetilde{n} (\theta') \rangle 
d\theta' = \arg \langle n(\theta_0)\ket{n(\theta)} ,  
\end{eqnarray}
by using Eq. (\ref{eq:mabpotential}). Clearly, $\gamma_{n}$ is locally
gauge invariant as it is independent of $\xi (\theta)$. It corresponds
to phase jumps of $\pi$ at points across which the real-valued
quantity $\langle n(\theta_0) \ket{n(\theta)}$ goes through zero and
changes sign. In the case where an even number of $\pi$ phase jumps
occurs, the nuclear factor $\chi(\theta)$ is single-valued and there
is no MAB effect. On the other hand, for an odd number of such jumps,
there is a physically nontrivial sign change for such a loop. Thus,
the presence of a nontrivial MAB effect could be explained locally as
it requires the existence of points along the nuclear path where the
electronic states at $\theta_{0}$ and $\theta$ become orthogonal. This
assignment of electronic Berry phase shift is gauge invariant at each
point along the nuclear path and thus experimentally testable in
principle. It shows that MAB does not obey the criterion (T2) for a
topological phase effect.

\section{Preferred gauge}
The assignment of a gauge invariant electronic Berry phase for any
open portion of the closed nuclear path suggests that MAB is not
topological and that it might be meaningful to introduce a notion of
preferred gauge in this context. The corresponding preferred vector
potential is defined as that whose line integral gives the open path
Berry phase. This idea can be put forward in terms of a so-called
gauge invariant reference section \cite{pati95a,pati95b} as follows.

First, in general terms, let $\ket{\psi (s)}$ be a normalized Hilbert 
space representative of the pure quantal state $\psi (s)$ tracing out 
the path ${\cal C}: s\in [s_0,s_1] \rightarrow \psi(s)$ in projective
Hilbert space ${\cal P}$. Now, if $0 \neq \big| \bra{\psi (s_0)} 
\psi (s_1)\rangle \big| \leq 1$, then the Berry phase $\gamma 
[{\cal C}]$ associated with ${\cal C}$ is \cite{mukunda93}
\begin{eqnarray} 
\gamma [{\cal C}] & = & \arg \bra{\psi (s_0)} \psi (s_1) \rangle - 
\int_{s_0}^{s_1} \arg \bra{\psi (s)} \psi (s+ds) \rangle 
\nonumber \\ 
 & = & \arg \bra{\psi (s_0)} \psi (s_1) \rangle + 
i \int_{s_0}^{s_1} \bra{\psi (s)} \dot{\psi} (s) \rangle \, ds , 
\label{eq:ncgp}
\end{eqnarray}
where the first and second term on the right-hand side contain 
the global phase and the accumulation of local phase changes,
respectively. This open path Berry phase is real-valued and
reparametrization invariant \cite{mukunda93}. It is gauge 
invariant and thereby measurable 
\cite{garcia98,wagh98,sjoqvist01,larsson03} in that it 
is independent of choice of Hilbert space representative.
$\gamma [{\cal C}]$ reduces to the cyclic Berry phase 
\cite{berry84} in the particular case where 
$\big| \bra{\psi (s_0)} \psi (s_1) \rangle \big| = 1$. 

A gauge invariant reference section $\ket{\phi (s)}$ is 
a nonlinear functional of $\ket{\psi (s)}$ defined as 
\cite{pati95a,pati95b} 
\begin{eqnarray}
\ket{\phi (s;s_0)} = 
\exp \Big( -i\arg \bra{\psi (s_0)} \psi (s) \rangle \Big) 
\ket{\psi (s)} ,
\label{eq:refsec} 
\end{eqnarray} 
which has the image ${\cal C}$ in ${\cal P}$. $\ket{\phi (s;s_0)}$ 
is in one to one correspondence with ${\cal C}$ in that it is gauge 
invariant under phase transformations of $\ket{\psi (s)}$. Furthermore, 
by inserting Eq.  (\ref{eq:refsec}) into Eq. (\ref{eq:ncgp}), we obtain
\begin{eqnarray}
\gamma [{\cal C}] = \int_{s_0}^{s_1} {\cal A} (s;s_0) ds ,  
\end{eqnarray}  
where the gauge function reads   
\begin{eqnarray} 
{\cal A} (s;s_0) = i \bra{\phi (s;s_0)} \partial_s \phi (s;s_0) \rangle . 
\label{eq:genrefsec}
\end{eqnarray} 

Now, the one to one correspondence between the path ${\cal C}$ 
and $\{ \ket{\phi_n (s;s_0)} |$ $s\in [s_0,s_1] \}$, and the
fact that the open path Berry phase can be directly expressed in terms
of the vector potential ${\cal A} (s;s_0)$, suggest that the gauge
invariant reference section defined in Eq. (\ref{eq:refsec}) has a
special status: it is natural to regard $\ket{\phi_n (s;s_0)}$
and ${\cal A} (s;s_0)$ to constitute a `preferred gauge' for the
measurable open path Berry phase. In doing so, one should take notice
that the functional form of the preferred vector potential ${\cal A}
(s;s_0)$ depends in general upon $\psi (s_0)$, but is otherwise
unique up to an additional term whose integral over the interval
$[s_0,s_1]$ equals an integral multiple of $2\pi$. In particular, once
$\psi (s_0)$ has been chosen the corresponding phase factor is
gauge invariant under phase transformations of $\ket{\psi (s)}$.

We next apply this idea and calculate the gauge invariant electronic
reference section relevant for MAB and its concomitant vector
potential in the case of real electronic Hamiltonians.  Let the
pseudorotational angle $\theta \in [\theta_0,\theta_0+2\pi]$
parametrize a closed path ${\cal C}_N$ in nuclear configuration
space. Let $\ket{n (\theta)}=R(\theta)\ket{n}$ be a Hilbert space
representative of a nondegenerate electronic energy eigenstate along
this path. Under these assumptions, consider a finite portion $\Delta
{\cal C}_N$ of ${\cal C}_N$. First, suppose that $\bra{n (\theta_0)}
n(\theta)
\rangle \neq 0, \ \theta \in \Delta {\cal C}_N$. Along such a 
$\Delta {\cal C}_N$, the gauge invariant reference section reads 
\begin{eqnarray}
\ket{\phi_n (\theta;\theta_0)} = 
\exp \Big( -i\arg \bra{n (\theta_0)} n(\theta) \rangle \Big) 
\ket{n(\theta)} = \pm \ket{n(\theta)} , 
\label{eq:nonode}
\end{eqnarray} 
where the sign is independent of $\theta$. This follows since 
$\bra{n (\theta_0)} n(\theta) \rangle$ is real-valued and may
only change sign where it vanishes. Inserting Eq. (\ref{eq:nonode}) 
into Eq. (\ref{eq:genrefsec}), we obtain 
${\cal A}_n (\theta;\theta_0) = 0, \ \theta \in \Delta {\cal C}_N$. 

Now, suppose there is a single isolated point $\theta_k \in \Delta
{\cal C}_N$ across which $\bra{n (\theta_0)} n(\theta_k) \rangle$ goes
through zero and changes sign. For such a point, it follows that
\begin{eqnarray}
\ket{\phi_n (\theta;\theta_0)} = \pm 
\exp \Big( i\pi h_s (\theta - \theta_k) \Big) \ket{n(\theta)} , \ 
\theta \in \Delta {\cal C}_N ,  
\end{eqnarray} 
where $h_s$ is the unit step function. This yields 
\begin{eqnarray}
{\cal A}_n (\theta;\theta_0) = -\pi 
\delta (\theta - \theta_k), \ \theta \in \Delta {\cal C}_N ,
\end{eqnarray} 
where we have used that $\partial_{\theta} h_s(\theta - \theta_k) = 
\delta (\theta - \theta_k)$. 

Extending this to the whole closed path ${\cal C}_N$, we may assume 
it contains $K$ angles $\theta_1, \ldots , \theta_K$, all across 
which $\bra{n(\theta_0)} n(\theta) \rangle$ goes through zero and 
changes sign. Then, the gauge invariant reference section reads
\begin{eqnarray}
\ket{\phi_n (\theta;\theta_0)} = 
\exp \left( i\pi \sum_{k=1}^{K} h_s (\theta - \theta_k)  \right)
\ket{n (\theta)}, \ \theta \in {\cal C}_N 
\end{eqnarray}
with the corresponding preferred vector potential   
\begin{eqnarray} 
{\cal A}_n (\theta;\theta_0) = -\pi \sum_{k=1}^{K} 
\delta (\theta - \theta_k) , \ \theta \in {\cal C}_N .
\label{eq:kpotential}  
\end{eqnarray}
Notice that ${\cal A}_n (\theta;\theta_0)$ is a local quantity 
as it only depends upon the angles $\theta_1, \ldots, \theta_K$, 
whose location is gauge invariant once the `initial' electronic 
eigenstate $n(\theta_0)$ has been chosen. 

There is a nontrivial MAB effect if and only if $K$ is odd, since
\begin{equation} 
\gamma_n [{\cal C}_N] = \oint_{{\cal C}_N} {\cal A}_n d\theta = -K\pi .
\end{equation}  
It is important to notice that this criterion for a nontrivial MAB is
only fulfilled if the effective nuclear motion has physical access to
the whole closed path ${\cal C}_N$. To see this, let $\langle n(\theta_0)
\ket{n(\theta)}$ change sign at the angles $\theta_1,\ldots,\theta_K$ 
along ${\cal C}_N$. In the preferred gauge the effective Hamiltonian 
operator for the nuclear motion in the $\theta$ direction reads  
\begin{eqnarray} 
H_{n} = 
\bra{\phi_n (\theta;\theta_0)} H \ket{\phi_n (\theta;\theta_0)} =   
\frac{1}{2} \Big[ p_{\theta} + {\cal A}_n (\theta;\theta_0) \Big]^2 
+ E_n (\theta) .  
\label{eq:prefham}
\end{eqnarray}
Suppose $E_n (\theta)$ comprises an infinite potential 
barrier over the angular range $\theta \in [\vartheta,\vartheta + 
\Delta \vartheta]$, $0<\Delta \vartheta < 2\pi$, creating an inaccessible 
part ${\cal C}_N (\Delta \vartheta)$ of ${\cal C}_N$ for the nuclear 
motion. Then, it is consistent with the boundary condition 
$\chi(\vartheta) = \chi (\vartheta + \Delta \vartheta) = 0$ to absorb 
the vector potential ${\cal A}_n$ into the phase of the nuclear factor. 

\begin{figure}[ht!] 
\begin{center} 
\includegraphics[width=8 cm]{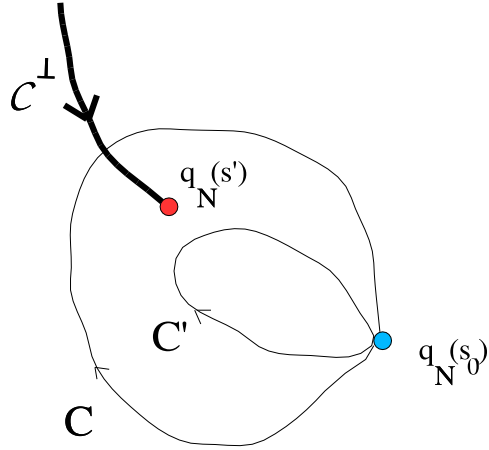} 
\end{center} 
\caption{Single line ${\cal C}^{\perp}: s \in [0,1] \rightarrow q_N (s)$ 
in nuclear configuration space for which ${\cal A} (q_N (s);q_N (s_0)) = 
-\pi \delta (q_N - q_N(s))$, for $0 \leq s < s' <1$, and ${\cal A} 
(q_N (s);q_N (s_0)) =0$, for $s' < s \leq 1$. For any closed path C that 
cross ${\cal C}^{\perp}$, the electronic eigenvector picks up a sign 
implying a degeneracy on any surface that has C as boundary. For any 
closed path C$'$ that does not cross ${\cal C}^{\perp}$, no such sign 
change occur. From Longuet-Higgins theorem \cite{longuet75} follows 
that $q_N(s')$ is a point of electronic degeneracy.}
\label{fig:degeneracy} 
\end{figure} 

Now, what is the origin of a nontrivial gauge invariant reference 
section? In part this can be answered in terms of the following
relationship between the gauge invariant reference section and
degeneracy points where two or more electronic potential energy
surfaces cross (for an analysis of the related connection between
degeneracies and singularities of the electronic eigenvectors, see
\cite{liyin90}). Let $q_N$ be the internal nuclear coordinates and
consider a line ${\cal C}^{\perp}: s \in [0,1] \rightarrow q_N (s)$ in
nuclear configuration space for which ${\cal A} (q_N (s);q_N (s_0)) =
-\pi \delta (q_N - q_N(s))$, for $0 \leq s < s' <1$, and ${\cal A}
(q_N (s);q_N (s_0)) =0$, for $s' < s \leq 1$, see
Fig. \ref{fig:degeneracy}. Now, any path C in the Born-Oppenheimer
regime that starts and ends at the reference point $q_N (s_0)$ and
cross ${\cal C}^{\perp}$ once, must be associated with a sign change
of the electronic eigenvector and thus must enclose at least one point
of degeneracy on any surface that has C as boundary, as implied by 
the Longuet-Higgins theorem \cite{longuet75}. On the other hand, 
there is no such sign change originating from ${\cal C}^{\perp}$ 
for a closed path C$'$ that does not cross ${\cal C}^{\perp}$, 
which implies that there must be one less degeneracy enclosed by 
such a path than by C. Thus, $q_N (s')$ is a degeneracy point.

To illustrate the gauge invariant reference section for MAB, let us
revisit the linear $+$ quadratic $E\otimes \varepsilon$ Jahn-Teller
effect, which is known to exhibit a nontrivial MAB structure. There,
the symmetry induced degeneracy of two electronic states ($E$) is
lifted by their interaction with a doubly degenerate vibrational mode
($\varepsilon$). In the vicinity of the degeneracy point at the symmetric
nuclear configuration, this may be modeled by the vibronic Hamiltonian
\cite{zwanziger87}
\begin{eqnarray}
H = \frac{1}{2} p_{r}^{2} + \frac{1}{2r^{2}} 
p_{\theta}^{2} + \frac{1}{2} r^{2} + \Delta {\cal E} (r,\theta) 
\big( \cos \alpha (r,\theta) \sigma_x + \sin \alpha (r,\theta) 
\sigma_z \big) . 
\label{eq:Exeham}
\end{eqnarray}
Here, $(r,\theta)$ are polar coordinates of the vibrational 
mode, $(p_{r},p_{\theta})$ the corresponding canonical 
momenta, $k\geq 0$ and $g\geq 0$ are the linear and quadratic 
vibronic coupling strength, respectively. $\Delta {\cal E}$ 
and $\alpha$ are given by  
\begin{eqnarray}
\Delta {\cal E} (r,\theta) & = & \sqrt{k^2r^2 + kgr^3\cos 3\theta + 
\frac{1}{4}g^2r^4} , 
\nonumber \\ 
\Delta {\cal E} (r,\theta) e^{i\alpha (r,\theta)} & = & kr \cos \theta + 
\frac{1}{2} gr^2 \cos 2\theta + i \big( kr \sin \theta - 
\frac{1}{2} gr^2 \sin 2\theta \big) . 
\end{eqnarray}
The electronic degrees of freedom are described by Pauli operators
defined in terms of the diabatic electronic states
$|0\rangle$ and $|1\rangle$ as $\sigma_{x} = |0 \rangle \langle 1| +
|1 \rangle \langle 0|$, $\sigma_{y} = -i|0 \rangle \langle 1| + i|1
\rangle \langle 0|$, and $\sigma_{z} = |0 \rangle \langle 0| - |1
\rangle \langle 1|$. Diagonalizing the electronic part of $H$ 
yields the electronic eigenvectors  
\begin{eqnarray}
\ket{+(\alpha)} & = & \cos \frac{\alpha}{2} \ket{0} + 
\sin \frac{\alpha}{2} \ket{1} ,
\nonumber \\ 
\ket{-(\alpha)} & = & -\sin \frac{\alpha}{2} \ket{0} + 
\cos \frac{\alpha}{2} \ket{1}   
\end{eqnarray}
with corresponding energies $E_{\pm} = \frac{1}{2} r^2 \pm \Delta 
{\cal E} (r,\theta)$ and where we have put $\alpha \equiv \alpha 
(r,\theta)$ for brevity. The phase $\alpha$ is undefined only if 
$\Delta {\cal E} (r,\theta) = 0$, corresponding to the degeneracy 
points at $r=0$ and $r=2k/g$ for $\theta=\pi/3,\pi,5\pi/3$.  

\begin{figure}[ht!] 
\begin{center} 
\includegraphics[width=8 cm]{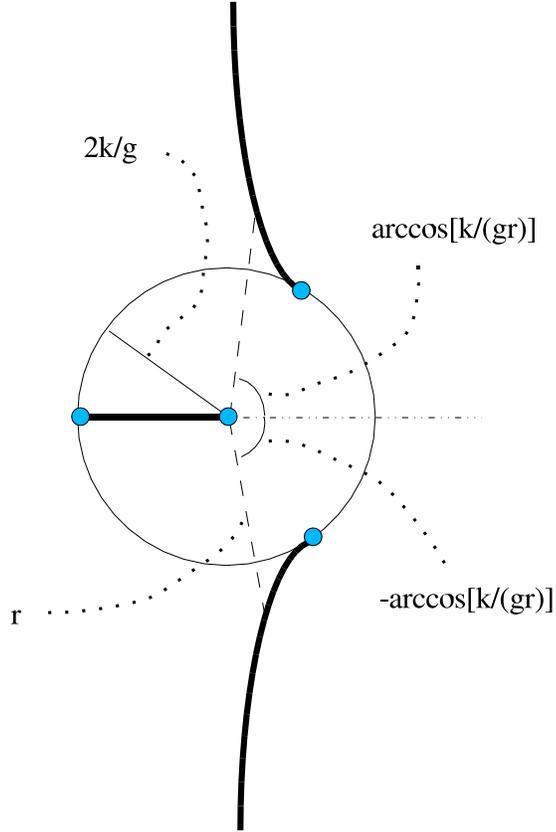} 
\end{center} 
\caption{Lines of sign change and degeneracy points in the linear $+$ 
quadratic $E\otimes \varepsilon$ Jahn-Teller model for any of the two 
Born-Oppenheimer states. Degeneracies are indicated by small circles; 
they occur at the origin and at $r = 2k/g$ for angles $\theta = 
\pi /3, \ \pi, \ 5\pi/3$. The electronic reference state is chosen 
at an arbitrary point along the dashed-dotted line ($\theta = 0$).
The thick continuous lines correspond to points where $\bra{\pm
(\alpha_0)} \pm (\alpha) \rangle$ goes through zero and changes sign.
These lines of sign change end either at infinity or at a degeneracy.
For a circular closed path with $r>2k/g$, there is an even number of
jumps for this path, and consequently no MAB effect for the
corresponding nuclear motion along the path. On the other hand, for
any circular closed path with $r < 2k/g$ there is a single node at
$\theta = \pi$, assigning a nontrivial MAB effect.}
\label{fig:nodes} 
\end{figure} 

To calculate the gauge invariant reference section, we may choose 
$\theta_0 =0$ so that $\alpha_0 \equiv \alpha (r_0,\theta_0) =0$, 
independent of $r_0$. In this case, we obtain 
\begin{eqnarray} 
\bra{\pm (\alpha_0)} \pm (\alpha) \rangle = 
\cos \frac{\alpha}{2} . 
\end{eqnarray} 
This vanishes for $\alpha = \pi$ corresponding to 
\begin{eqnarray}
kr \sin \theta - \frac{1}{2} gr^2 \sin 2\theta & = & 0 , 
\nonumber \\ 
kr \cos \theta + \frac{1}{2} gr^2 \cos 2\theta & < & 0 ,
\end{eqnarray}
which have the solutions $\theta' = \pi$ for $r < 2k/g$ and $\theta' =
\pm \arccos [k/(gr)]$ for $r>2k/g$. Since the solutions for $r<2k/g$ 
are independent of $r$, they constitute a radial line whose
end-points are the electronic degeneracies at the origin and at
$(r,\theta') = (2k/g,\pi)$, see Fig. \ref{fig:nodes}. If $2k/g <r 
\rightarrow \infty$, we obtain the limit angles $\theta' = \pm \pi /2$ 
at the infinity. On the other hand, if $r \rightarrow 2k/g^{+}$, 
then the lines of sign change terminate at $\theta' = \pm \pi /3$, 
which are the remaining electronic degeneracies, see Fig. \ref{fig:nodes}.    

Next, let us consider the gauge invariant reference section and 
the corresponding vector potential. For $r<2k/g$, we have  
\begin{eqnarray}
\ket{\phi_{\pm}} & = & 
\exp \Big( i\pi h_s (\theta -\pi) \Big) \ket{\pm (\alpha)} , 
\nonumber \\  
{\cal A}_{\pm} (\theta;0) & = & -\pi \delta (\theta - \pi) , 
\end{eqnarray}
while for $r>2k/g$, we have
\begin{eqnarray}
\ket{\phi_{\pm}} & = & 
\exp \Big( i\pi h_s (\theta -\arccos [k/(gr)]) 
\nonumber \\ 
 & & + 
i\pi h_s (\theta +\arccos [k/(gr)]) \Big) 
\ket{\pm (\alpha)} , 
\nonumber \\  
{\cal A}_{\pm} (\theta;0) & = & 
-\pi \delta (\theta -\arccos [k/(gr)]) 
-\pi \delta (\theta +\arccos [k/(gr)]).  
\end{eqnarray}

\section{Aharonov-Casher analogue} 
It was demonstrated in \cite{sjoqvist02} that for the linear or 
the quadratic $E\otimes \varepsilon$ Jahn-Teller system, there is a 
close analogy with the Aharonov-Casher (AC) phase effect \cite{aharonov84}  
for an electrically neutral spin$-\frac{1}{2}$ particle encircling a 
line of charge. From this perspective, one may regard the presence 
(absence) of MAB effect in the linear (quadratic) case as a nontrivial 
(trivial) $\pi$ ($2\pi$) AC phase shift. Here, we extend this idea to 
the linear $+$ quadratic case. 

In brief, the AC effect may occur when an electrically 
neutral spin$-\frac{1}{2}$ particle carrying a magnetic dipole moment 
$\mu$ encircles a straight line of charge. Under the condition that 
the spin is parallel to the charged line, the particle acquires along 
the closed path $\partial S$ the phase shift ($\hbar = 1$) 
\begin{eqnarray}
\gamma_{{\textrm{\footnotesize AC}}} = 
\frac{\mu}{c^2} \oint_{\partial S} 
\big( \hat{{\bf n}} \times {\bf E} \big) \cdot d{\bf r} = 
\frac{\mu}{c^2} \int \! \! \! \int_S \nabla \cdot {\bf E} \ dS = 
\frac{\mu \lambda}{c^2\epsilon_0}  
\end{eqnarray} 
with $c$ the speed of light, $\hat{{\bf n}}$ the direction of the
dipole, ${\bf E}$ the electric field, $d{\bf r}$ a line element along
this path, $S$ any surface with $\partial S$ as boundary, $\lambda$
the enclosed charge density, and $\epsilon_0$ the electric vacuum
permittivity. 

To demonstrate the promised analogy between the MAB effect in the
$E\otimes \varepsilon$ Jahn-Teller system and the AC effect, we may
use the apparent spin analogy in terms of which we may notice that the
electronic Hamiltonian in Eq. (\ref{eq:Exeham}) describes a
spin$-\frac{1}{2}$ under influence of an effective magnetic field that
rotates around the $\sigma_y$ axis by the angle $\alpha
(r,\theta)$. We expect the electronic Hamiltonian be fixed (possibly
up to an unimportant $r$ and $\theta$ dependent scale factor) in an
internal molecular frame that co-moves with this rotation. The
essential point here is that, contrary to the usual case of a
spin$-\frac{1}{2}$ in a rotating external magnetic field, the rotation
angle $\alpha$ of this Jahn-Teller system depends upon the internal
variables $r$ and $\theta$.  This has the consequence that the
vibronic Hamiltonian in the co-moving frame, which reads (omitting the
inessential $\frac{1}{2}p_r^2$ and $\frac{1}{2} r^2$ terms)
\begin{eqnarray} 
H' = U^{\dagger} H U = 
\frac{1}{2r^{2}} \left[ p_{\theta} - \frac{1}{2} \partial_{\theta} 
\alpha (r,\theta) \sigma_{y} \right]^{2} + 
\Delta {\cal E} (r,\theta) \sigma_{z} 
\end{eqnarray} 
with $U = \exp [ -i \alpha (r,\theta) \sigma_{y}/2 ]$ the unitary 
spin rotation operator, contains a nontrivial modification of the 
nuclear kinetic energy operator. Indeed, by using $H'$ and the 
Heisenberg picture, this modification in turn affects the equations 
of motion for the electronic variables, which read  
\begin{eqnarray} 
\dot{\sigma}_{x} & = & -\partial_{\theta} \alpha (r,\theta) 
\dot{\theta} \ \sigma_z - 2\Delta {\cal E} (r,\theta) \ \sigma_y , 
\nonumber \\   
\dot{\sigma}_{y} & = & 2\Delta {\cal E} (r,\theta) \ \sigma_x , 
\nonumber \\  
\dot{\sigma}_{z} & = & \partial_{\theta} \alpha (r,\theta) 
\dot{\theta} \ \sigma_x , 
\label{eq:eqm}
\end{eqnarray} 
where $r^{2} \dot{\theta} = p_{\theta} - \frac{1}{2} 
\partial_{\theta} \alpha (r,\theta) \sigma_{y}$. The 
Born-Oppenheimer regime is characterized by the condition 
$\partial_{\theta} \alpha (r,\theta) |\dot{\theta}| \ll \Delta 
{\cal E}(r,\theta)$ that apparently breaks down when 
$\Delta {\cal E}(r,\theta)$ is very small, which happens close to 
the electronic degeneracies. From Eq. (\ref{eq:eqm}), it follows 
that the electronic motion describes the local torque due to an 
effective magnetic field ${\bf B}_{\textrm{\footnotesize eff}} = 
-\partial_{\theta} \alpha (r,\theta) \dot{\theta} \, {\bf e}_y + 
2\Delta {\cal E} (r,\theta) \, {\bf e}_z$ seen by the electronic 
variables in the rotating frame. The large $z$ component of 
${\bf B}_{\textrm{\footnotesize eff}}$ depends only on the energy 
difference $\Delta {\cal E}(r,\theta)$ between the two electronic 
states and is thus irrelevant to MAB. On the other hand, the small 
$y$ component corresponds exactly to the MAB effect and gives rise 
to a ${\mbox{\boldmath $\sigma$}} \cdot ({\bf v} \times 
{\bf E}_{\textrm{\footnotesize eff}})$ term for the 
electronic variables in the co-moving frame in the $x-z$ 
plane. Explicitly, we have   
\begin{equation} 
{\bf E}_{\textrm{\footnotesize eff}} = 
\frac{\partial_{\theta} \alpha (r,\theta)}{2r} \ {\bf e}_{r} + 
E_{\theta} (r,\theta) \, {\bf e}_{\theta} , 
\end{equation}
where we have left out the explicit form of the $\theta$ component of
${\bf E}_{\textrm{\footnotesize eff}}$, as it does not contribute to
the MAB phase effect. The effect of this field is equivalent to that
of an AC system that consists of four charged lines in the $y$
direction sitting at the four conical intersections at $r=0$ and
$r=2k/g$ for $\theta = \pi/3 , \pi , 5\pi/3$, all of which with the
charge per unit length being proportional to $\frac{1}{2}$. Thus, the
MAB effect for the $E\otimes \varepsilon$ Jahn-Teller system resembles
exactly that of the AC effect for an electrically neutral
spin$-\frac{1}{2}$ particle encircling a certain configuration of
charged lines perpendicular to the plane of motion.

Note that the phase shift $\gamma_{{\textrm{\footnotesize AC}}}$
only depends upon the enclosed charge, but is independent of the shape
of the dipole's path. On the other hand, the dipole feels the gauge
invariant electric field ${\bf E}$, which defines a preferred gauge in
terms of the gauge invariant effective vector potential $(\mu /c^2)
\hat{{\bf n}} \times {\bf E}$ and which causes a nontrivial, essentially 
local and nontopological autocorrelation among the spin variables, as
demonstrated in \cite{peshkin95}. From this perspective, we believe 
that the present analogy between the AC and MAB effects further 
strengthens the local and nontopological interpretation of MAB 
for this Jahn-Teller system.

\section{Conclusions}
Arguably one of the most intriguing discoveries of the second 
half of the 20$th$ century was that of a nonlocal and topological 
interference effect for a charged particle moving around a magnetic 
flux line, made by Aharonov and Bohm (AB) \cite{aharonov59}. 
Analogues of this remarkable effect have since then been found, 
such as the Aharonov-Casher (AC) effect \cite{aharonov84} and 
its Maxwell  dual, the so-called He-McKellar-Wilkens (HMW) effect 
\cite{he93,wilkens94}. Common to these analogue effects are that 
they only occur under certain restrictions on some additional degree 
of freedom: the spin direction for AC and the direction of an electric 
dipole for HMW. These additional restrictions turn out to make these 
effects essentially different from the standard AB, in that they do 
not obey all the properties for being nonlocal and topological. 

Does the molecular Aharonov-Bohm (MAB) effect share the fate of these
other analogue effects? This question was examined quite recently by
the present author \cite{sjoqvist02}, in the particular case of the
$E\otimes \varepsilon$ Jahn-Teller system. In identifying the
restrictions for this system, which are the conditions of attaining
the Born-Oppenheimer regime and the electronic state being in one of
two instantaneous energy eigenstates, this issue can be addressed very
much along the line of the other analogue effects. The outcome was:
the MAB effect for this system is neither nonlocal nor topological in
the sense of the standard AB effect. In this paper, this result has
been generalized to any molecular system with real electronic
Hamiltonian and where a nontrivial MAB effect shows up. In addition, a
notion of preferred gauge for such systems, defined by an effective
vector potential whose open path integral is the open path Berry phase
for the electronic motion, has been suggested.

One of the early motivations preceding the analysis in 
\cite{sjoqvist02} was to examine whether there is a relation 
between the MAB effect in the $E\otimes \varepsilon$ Jahn-Teller
system and the AC effect, based upon the simple observation that the
original vibronic Hamiltonian for this Jahn-Teller system resembles
exactly that of an electrically neutral spin$-\frac{1}{2}$ in a
certain field configuration. Indeed, by transforming to a molecular
frame that co-moves with the nuclear pseudorotation, this analogue was
made explicit in \cite{sjoqvist02} in the case of either linear or
quadratic coupling. In the present paper, this result has been
extended to the linear $+$ quadratic case, leading to an anisotropic
effective electric field originating from four charged lines sitting
at the four conical intersections in nuclear configuration space.

\section*{Acknowledgments} 
I wish to express my deep gratitude to Prof. Osvaldo Goscinski,
especially for introducing me to the subject of geometric phases and
how they appear in molecular systems, but also for numerous
discussions and collaboration during the past 10 years. Osvaldo's
great impact on my scientific thinking makes it a very special honour
to dedicate this paper on his 65$th$ birthday. I also wish to thank
Mauritz Andersson, Henrik Carlsen, Marie Ericsson, Gonzalo Garc\'{\i}a
de Polavieja, Magnus Hedstr\"om, and Niklas Johansson for discussions
and collaboration over the years on issues related to the present
paper. This work was supported by the Swedish Research Council.


\begin{thebibliography}{99}  
\bibitem{longuet58} H.C. Longuet-Higgins, U. \"{O}pik, 
M.H.L. Pryce and R.A. Sack, 
Proc. Roy. Soc. London Ser. A {\bf 244}, 1 (1958).  
\bibitem{longuet61} H.C. Longuet-Higgins, 
Adv. Spectr. {\bf 2}, 429 (1961). 
\bibitem{herzberg63} G. Herzberg and H.C. Longuet-Higgins, 
Disc. Frad. Soc. {\bf 35}, 77 (1963). 
\bibitem{longuet75} H.C. Longuet-Higgins, 
Proc. Roy. Soc. London Ser. A {\bf 344}, 147 (1975). 
\bibitem{berry84} M.V. Berry, 
Proc. Roy. Soc. London Ser. A {\bf 392}, 45 (1984).
\bibitem{kendrick97} B. Kendrick, 
Phys. Rev. Lett. {\bf 79}, 2431 (1997).  
\bibitem{vonbusch98} H. von Busch, V. Dev, H.-A. Eckel, S. Kasahara, 
J. Wang, W. Demtr\"{o}der, P. Sebald, and W. Meyer, 
Phys. Rev. Lett. {\bf 81}, 4584 (1998).  
\bibitem{kupperman93} A. Kupperman and Y.M. Wu, 
Chem. Phys. Lett. {\bf 205}, 577 (1993).
\bibitem{kendrick96} B. Kendrick and R.T. Pack, 
J. Chem. Phys. {\bf 104}, 7475 (1996).  
\bibitem{adhikari00} S. Adhikari and G.D. Billing, 
Chem. Phys. {\bf 259}, 149 (2000).  
\bibitem{sjoqvist94} E. Sj\"{o}qvist and O. Goscinski, 
Chem. Phys. {\bf 186}, 17 (1994). 
\bibitem{ham87} F.S. Ham,
Phys. Rev. Lett. {\bf 58}, 725 (1987). 
\bibitem{ham90} F.S. Ham,
J. Phys.: Condens. Matter {\bf 2}, 1163 (1990). 
\bibitem{rios96} P. De Los Rios, N. Manini, and E. Tosatti, 
Phys. Rev. B  {\bf 54}, 7157 (1996). 
\bibitem{moate96} C.P. Moate, M.C.M. O'Brien, J.L. Dunn, C.A. Bates, 
Y.M. Liu, and V.Z. Polinger, 
Phys. Rev. Lett. {\bf 77}, 4362 (1996). 
\bibitem{stone76} A.J. Stone, 
Proc. Roy. Soc. London Ser. A {\bf 351}, 141 (1976). 
\bibitem{varandas79} A.J.C. Varandas, J. Tennyson, and J.N. Murrell, 
Chem. Phys. Lett. {\bf 61}, 431 (1979). 
\bibitem{xantheas90} S. Xantheas, S.T. Elbert, and K. Ruedenberg, 
J. Chem. Phys. {\bf 93}, 7519 (1990).  
\bibitem{ceotto00} M. Ceotto and F.A. Gianturco, 
J. Chem. Phys. {\bf 112}, 5820 (2000). 
\bibitem{johansson03} N. Johansson and E. Sj\"{o}qvist, 
Phys. Rev. Lett. {\bf 92}, 060406 (2004). 
\bibitem{mead79} C.A. Mead and D.G. Truhlar, 
J. Chem. Phys. {\bf 70}, 2284 (1979); 
\bibitem{mead80} C.A. Mead, 
Chem. Phys. {\bf 49}, 23 (1980); {\it Ibid.} {\bf 49}, 33 (1980).  
\bibitem{aharonov59} Y. Aharonov and D. Bohm, 
Phys. Rev. {\bf 115}, 485 (1959). 
\bibitem{sjoqvist02} E. Sj\"{o}qvist,
Phys. Rev. Lett. {\bf 89}, 210401 (2002). 
\bibitem{pati95a} A.K. Pati, 
J. Phys. A {\bf 28}, 2087 (1995).  
\bibitem{pati95b} A.K. Pati, 
Phys. Rev. A {\bf 52}, 2576 (1995). 
\bibitem{samuel88} J. Samuel and R. Bhandari, 
Phys. Rev. Lett. {\bf 60}, 2339 (1988). 
\bibitem{mukunda93} N. Mukunda and R. Simon, 
Ann. Phys. (N.Y.) {\bf 288}, 205 (1993).  
\bibitem{peshkin95} M. Peshkin and H.J. Lipkin, 
Phys. Rev. Lett. {\bf 74}, 2847 (1995). 
\bibitem{sjoqvist97} E. Sj\"{o}qvist and M. Hedstr\"{o}m, 
Phys. Rev. A {\bf 56}, 3417 (1997).
\bibitem{garcia98} G. Garc\'{\i}a de Polavieja and E. Sj\"{o}qvist, 
Am. J. Phys. {\bf 66}, 431 (1998).  
\bibitem{englman99} R. Englman and A. Yahalom, 
Phys. Rev. A {\bf 60}, 1802 (1999); 
\bibitem{englman00} R. Englman, A. Yahalom, and M. Baer, 
Eur. Phys. J. D {\bf 8}, 1 (2000). 
\bibitem{aharonov84} Y. Aharonov and A. Casher, 
Phys. Rev. Lett. {\bf 53}, 319 (1984). 
\bibitem{larsson03} P. Larsson and E. Sj\"{o}qvist, 
Phys. Rev. A {\bf 68}, 042109 (2003).  
\bibitem{wagh98} A.G. Wagh, V.C. Rakhecha, P. Fischer, and A. Ioffe, 
Phys. Rev. Lett. {\bf 81}, 1992 (1998). 
\bibitem{sjoqvist01} E. Sj\"{o}qvist, 
Phys. Lett. A {\bf 286}, 4 (2001). 
\bibitem{liyin90} L. Yin and O. Goscinski, 
Int. J. Quantum Chem. {\bf 37}, 249 (1990). 
\bibitem{zwanziger87} J.W. Zwanziger and E.R. Grant, 
J. Chem. Phys. {\bf 87}, 2954 (1987). 
\bibitem{he93} X.G. He and B.H.J. McKellar, 
Phys. Rev. A {\bf 47}, 3424 (1993).   
\bibitem{wilkens94} M. Wilkens, 
Phys. Rev. Lett. {\bf 72}, 5 (1994).   
\end{thebibliography}
\end{document}